\documentclass[seceq,letter,twocolumn]{jpsj3}

\usepackage{color}

\def\runauthor{
M. Matsumoto, M. Koga, and H. Kusunose
}

\title{
Coexistence of Even- and Odd-Frequency Superconductivities \\
Under Broken Time-Reversal Symmetry
}

\author{
Masashige Matsumoto\thanks{E-mail address: spmmatu@ipc.shizuoka.ac.jp}, Mikito Koga$^1$, and Hiroaki Kusunose$^2$
}

\inst{
Department of Physics, Faculty of Science, Shizuoka University, Shizuoka 422-8529, Japan \\
$^1$Department of Physics, Faculty of Education, Shizuoka University, Shizuoka 422-8529, Japan \\
$^2$Department of Physics, Ehime University, Matsuyama, Ehime 790-8577, Japan
}

\recdate{December 16, 2011}

\abst{
A novel superconducting state under the broken time-reversal symmetry is studied
in conventional phonon-mediated superconductors.
By solving the Eliashberg equation self-consistently with the mass renormalization effect,
it is found that the even- and odd-frequency components of the order parameter coexist in the bulk system
as a consequence of the broken time-reversal symmetry.
This finding would direct more attention to the odd-frequency pairing
that affects physical quantities, especially in strong coupling superconductors.
}

\kword{
odd-frequency superconductivity, broken time-reversal symmetry, spin-triplet $s$-wave state
}

\begin{document}

\maketitle

\newcommand{\ri}{{\rm i}}
\renewcommand{\H}{{\mathcal H}}
\newcommand{\bS}{{\mbox{\boldmath$S$}}}

\newcommand{\bd}{{\mbox{\boldmath$d$}}}

\newcommand{\brho}{{\mbox{\boldmath$\rho$}}}

\newcommand{\bsigma}{{\mbox{\boldmath$\sigma$}}}

\newcommand{\bSigma}{{\mbox{\boldmath$\Sigma$}}}
\newcommand{\bSigmap}{{\mbox{\boldmath$\Sigma$}}_{\rm p}}
\newcommand{\bSigmai}{{\mbox{\boldmath$\Sigma$}}_{\rm i}}
\newcommand{\bSigmad}{{\mbox{\boldmath$\Sigma$}}_{\Delta}}

\newcommand{\bDelta}{{\mbox{\boldmath$\Delta$}}}

\newcommand{\bpsi}{{\mbox{\boldmath$\psi$}}}
\newcommand{\bG}{{\mbox{\boldmath$G$}}}
\newcommand{\bT}{{\mbox{\boldmath$T$}}}
\newcommand{\bQ}{{\mbox{\boldmath$Q$}}}
\newcommand{\br}{{\mbox{\boldmath$r$}}}
\newcommand{\bH}{{\mbox{\boldmath$H$}}}
\newcommand{\bR}{{\mbox{\boldmath$R$}}}
\newcommand{\bq}{{\mbox{\boldmath$q$}}}
\newcommand{\bu}{{\mbox{\boldmath$u$}}}
\newcommand{\bv}{{\mbox{\boldmath$v$}}}
\newcommand{\bk}{{\mbox{\boldmath$k$}}}
\newcommand{\balpha}{{\mbox{\boldmath$\alpha$}}}
\newcommand{\ba}{{\mbox{\boldmath$a$}}}
\newcommand{\bskp}{{\mbox{\scriptsize \boldmath$k$}}}
\newcommand{\bsrp}{{\mbox{\scriptsize \boldmath$r$}}}
\newcommand{\bsRp}{{\mbox{\scriptsize \boldmath$R$}}}
\newcommand{\bsQp}{{\mbox{\scriptsize \boldmath$Q$}}}
\newcommand{\bsqp}{{\mbox{\scriptsize \boldmath$q$}}}
\newcommand{\bsk}{\bskp}
\newcommand{\bsr}{\bsrp}
\newcommand{\bsR}{\bsRp}
\newcommand{\bsQ}{\bsQp}
\newcommand{\bsq}{\bsqp}
\newcommand{\re}{{\rm e}}
\newcommand{\rd}{{\rm d}}
\newcommand{\va}{{\vec{a}}}
\newcommand{\vX}{{\vec{X}}}
\newcommand{\hepsilon}{{\hat{\epsilon}}}
\newcommand{\hX}{{\hat{X}}}
\newcommand{\sqp}{{\mbox{\scriptsize $q$}}}
\newcommand{\sq}{\sqp}
\newcommand{\sxp}{{\mbox{\scriptsize $x$}}}
\newcommand{\sx}{\sxp}
\newcommand{\syp}{{\mbox{\scriptsize $y$}}}
\newcommand{\sy}{\syp}
\newcommand{\szp}{{\mbox{\scriptsize $z$}}}
\newcommand{\sz}{\szp}

\newcommand{\tomega}{\tilde{\omega}}

\newcommand{\tth}{\tilde{h}}

\newcommand{\omegae}{\omega_{\rm E}}

\newcommand{\Sigmap}{\Sigma_{\rm p}}
\newcommand{\Sigmai}{\Sigma_{\rm i}}
\newcommand{\Sigmah}{\Sigma_h}
\newcommand{\Sigmae}{\Sigma_{\rm e}}
\newcommand{\Sigmao}{\Sigma_{\rm o}}

\newcommand{\Deltauu}{\Delta_{\uparrow\uparrow}}

\newcommand{\Deltaa}{\Delta^{(1)}_{\uparrow\uparrow}}
\newcommand{\Deltab}{\Delta^{(1)}_{\uparrow\downarrow}}
\newcommand{\Deltac}{\Delta^{(1)}_{\downarrow\uparrow}}
\newcommand{\Deltad}{\Delta^{(1)}_{\downarrow\downarrow}}
\newcommand{\Deltae}{\Delta^{(2)}_{\uparrow\uparrow}}
\newcommand{\Deltaf}{\Delta^{(2)}_{\uparrow\downarrow}}
\newcommand{\Deltag}{\Delta^{(2)}_{\downarrow\uparrow}}
\newcommand{\Deltah}{\Delta^{(2)}_{\downarrow\downarrow}}
\newcommand{\Deltaea}{\Delta_{\rm e}^{(1)}}
\newcommand{\Deltaoa}{\Delta_{\rm o}^{(1)}}
\newcommand{\Deltaeb}{\Delta_{\rm e}^{(2)}}
\newcommand{\Deltaob}{\Delta_{\rm o}^{(2)}}

\newcommand{\Deltaev}{\Delta^{\rm e}}
\newcommand{\Deltaod}{\Delta^{\rm o}}

\newcommand{\Tc}{T_{\rm c}}


Since superconductivity was discovered by Kamerlingh Onnes about a century ago,
it has been one of the most important subjects in the condensed matter physics.
Odd-frequency superconductivity is one of the issues attracting interest in this field today.
It was originally suggested by Berezinskii in the $^3$He superfluid phase
\cite{Berezinskii}
and was developed further by Kirkpatrick and Belitz
\cite{Kirkpatrick,Belitz}
and by Balatsky and co-workers
\cite{Balatsky,Abrahams-1,Abrahams-2}
in conduction electron systems.
The possibility of the odd-frequency superconductivity was also discussed in doped triangular antiferromagnets
\cite{Vojta}
and in electron systems with critical spin fluctuations in the vicinity of the quantum critical point.
\cite{Fuseya}
In a similar context, quasi-one-dimensional organic systems were also studied.
\cite{Shigeta-1,Shigeta-2}

In these works, the random phase approximation was frequently applied
to examine the stability of the odd-frequency superconductivity,
where the mass renormalization effect was neglected completely.
However, it was pointed out that mass renormalization tends to destabilize the odd-frequency superconductivity.
\cite{Abrahams-1}
Thus, a more elaborate discussion on the realization of the odd-frequency superconductivity
with the appropriate mass renormalization and vertex corrections is required.

In this respect, the odd-frequency superconductivity is not so easy to emerge by itself.
Alternatively, it is much easier to emerge or to be induced with the help of even-frequency superconductivity.
This is analogous to the parity mixing of Cooper pairs, such as $s$+$p$-wave, under the broken inversion symmetry.
\cite{Gorkov}
Similarly, the even- and odd-frequency components of the order parameter should coexist under the broken time-reversal symmetry.
The possibility of such an induced odd-frequency superconductivity has also been examined in symmetry reduction at interfaces
\cite{Bergeret,Linder-2009,Linder-2010,Yokoyama,Tanaka}
or vortices,
\cite{Tanuma}
and in fully spin-polarized electron systems with orbital fluctuations.
\cite{Hotta}
Since the previous studies did not examine the self-consistent solution of the order parameters in the superconducting phase,
there is still room to reexamine the induced odd-frequency superconductivity.

In this Letter, we explore the coexistence of the even- and odd-frequency pairings
in conventional phonon-mediated bulk superconductors on the basis of the Eliashberg theory,
where mass renormalization is appropriately taken into account.
\cite{Eliashberg}
We also emphasize the importance of an explicit self-consistent procedure
of the Eliashberg equation in the presence of the odd-frequency pairing.

As the simplest source breaking the time-reversal symmetry,
a uniform external magnetic field coupled to conduction electron spins is introduced here.
We note that other magnetic sources,
e.g., ferromagnets or ferromagnetically polarized impurities,
give essentially the same consequences.
To stabilize superconductivity, we consider the electron-phonon interaction as usual.
To make the discussion clearer, we focus on an Einstein phonon characterized by a single excitation energy.
The electron-phonon interaction gives rise to attractive interactions
for both spin-singlet and spin-triplet pairing channels.
\cite{Kusunose-odd}
The local nature of the Einstein phonon favors the $s$-wave pairing.
Since the triplet $s$-wave pairing is forbidden with the even-frequency dependence owing to the fermion property,
we must extend the theoretical framework including the odd-frequency dependence.
\cite{Berezinskii,Kusunose-odd}

We treat the electron-phonon interaction as the perturbation and
derive the lowest self-energy in $4\times 4$ matrix form as follows:
\cite{Eliashberg,Kusunose-odd}
\begin{align}
\bSigma(\ri\omega_l) = - g^2 \frac{N_{\rm p}}{\Omega} T \sum_m D(\ri\omega_l-\ri\omega_m)
\sum_\bsk \brho_3 \bG(\ri\omega_m,\bk) \brho_3.
\label{eqn:self-energy}
\end{align}
Here, $g$ represents the coupling constant for the electron-phonon interaction.
$N_{\rm p}$ is the number of phonon sites.
$\Omega$ represents the volume of the system.
$D(\ri\omega_l-\ri\omega_m)$ and $\bG(\ri\omega_m,\bk)$ are Green's functions for the phonon and conduction electron:
\begin{align}
&D(\ri\nu_l) = \frac{1}{ \ri\nu_l - \omegae} + \frac{1}{- \ri\nu_l - \omegae}, \cr
&\bG(\ri\omega_l,\bk) = \left[ \ri\omega_l - \epsilon_\bsk \brho_3 - h \brho_3 \bsigma_3 - \bSigma(\ri\omega_l) \right]^{-1}.
\label{eqn:G}
\end{align}
Here, $\omega_{\rm E}$ represents the phonon frequency.
$\epsilon_\bsk$ is the kinetic energy of the conduction electron measured from the Fermi energy.
$h$ represents the external magnetic field along the $z$-axis.
$\nu_l=2\pi T l$ and $\omega_l=\pi T(2l+1)$ are Matsubara frequencies at the temperature $T$ for bosons and fermions, respectively.
$\brho_\alpha$ and $\bsigma_\alpha$ $(\alpha=1,2,3)$ are Pauli matrices for the particle-hole and spin spaces, respectively.
The self-energy $\bSigma(\ri\omega_l)$ in eq. (\ref{eqn:G}) is given by
\begin{align}
\bSigma(\ri\omega_l) = - \ri \Sigma^\omega_l + \Sigma^h_l \brho_3 \bsigma_3 + \bSigma_\Delta(\ri\omega_l).
\label{eqn:sigma}
\end{align}
The first and second terms are for the normal parts that renormalize the Matsubara frequency and magnetic field, respectively.
There is no $\brho_3$ term in eq. (\ref{eqn:sigma}),
since it disappears after the summation over $\bk$ in eq. (\ref{eqn:self-energy}) under the particle-hole symmetry.
The third term in eq. (\ref{eqn:sigma}) is the anomalous part.

To consider the anomalous part, let us discuss what types of matrices come out from $\bG(\ri\omega_l,\bk)$ in eq. (\ref{eqn:G}).
Since the singlet pairing ($\brho_2 \bsigma_2$)
\cite{Rickayzen}
is stabilized at low temperatures, we assume that it is finite.
Then, the product of the singlet component $(\brho_2\bsigma_2$) and the Zeeman splitting term $(\brho_3\bsigma_3$)
gives rise to the following additional matrix component:
$\brho_3\bsigma_3\brho_2\bsigma_2 \propto \brho_1\bsigma_1$.
Moreover, a finite $\brho_1\bsigma_1$ component results in
$\brho_3\bsigma_3\brho_1\bsigma_1 \propto \brho_2\bsigma_2$.
The $\brho_1\bsigma_1$ component represents the $S_z=0$ spin-triplet pairing.
Since the orbital of the Cooper pair is of the $s$-wave type,
the $\brho_1 \bsigma_1$ component must exhibit the odd-frequency dependence.
\cite{Berezinskii}
Therefore, the anomalous self-energy should have the following form:
\begin{align}
\bSigma_\Delta(\ri\omega_l) = \Deltaev_l \brho_2 \bsigma_2 + \ri \Deltaod_l \brho_1 \bsigma_1.
\label{eqn:sigma-delta}
\end{align}
Here, the first term $\Deltaev_l \brho_2 \bsigma_2$ represents the $s$-wave singlet pairing with the even-frequency dependence.
The second term $\ri \Deltaod_l \brho_1 \bsigma_1$ is for the $s$-wave triplet pairing with the odd-frequency dependence.
With this choice of the relative phase between the even- and odd-frequency components,
$\Deltaev_l$ and $\Deltaod_l$ can be chosen as real quantities in the Eliashberg equation, as shown later.

In the presence of the magnetic field, the $S_z$ value along the field is a good quantum number.
However, the total $S$ is not the case.
Therefore, the singlet and $S_z=0$ triplet components are mixed,
while the other $S_z=\pm 1$ triplet components are decoupled completely from the singlet component.
\cite{equal-spin}
Thus, the $s$-wave triplet with the odd-frequency dependence appears inevitably with the singlet under the magnetic field.
In fact, this type of self-energy was taken into account years ago
to study the pair breaking effect caused by magnetic fields or magnetic impurities.
\cite{Fulde}
At that time, however, the self-consistent treatment for $\Deltaod_l$ was not performed,
since it was premature to consider the odd-frequency superconductivity.

In this work, we solve the Eliashberg equation self-consistently and reveal
that the odd-frequency component appears simultaneously with the even-frequency one
below the superconducting transition temperature $\Tc$.
Substituting eq. (\ref{eqn:G}) into eq. (\ref{eqn:self-energy}),
we obtain the Eliashberg equation in the following matrix form:
\begin{subequations}
\begin{align}
&\left(
  \begin{array}{c}
    \Sigma^\omega_l \cr
    \Sigma^h_l
  \end{array}
\right)
= \lambda T \sum_{m\ge 0}
\left(
  \begin{array}{cc}
    V_{lm}^- f^+_m & - V_{lm}^- f^-_m \cr
    V_{lm}^+ f^-_m &   V_{lm}^+ f^+_m
  \end{array}
\right)
\left(
  \begin{array}{c}
    \tomega_m \cr
    \tth_m
  \end{array}
\right),
\label{eqn:gap1} \\
&\left(
  \begin{array}{c}
    \Deltaev_l \cr
    \Deltaod_l
  \end{array}
\right)
= \lambda T \sum_{m\ge 0}
\left(
  \begin{array}{cc}
    V_{lm}^+ f^+_m & V_{lm}^+ f^-_m \cr
  - V_{lm}^- f^-_m & V_{lm}^- f^+_m
  \end{array}
\right)
\left(
  \begin{array}{c}
    \Deltaev_m \cr
    \Deltaod_m
  \end{array}
\right).
\label{eqn:gap2}
\end{align}
\label{eqn:gap}
\end{subequations}
Here, $\lambda = 2 g^2 N_{\rm p} N(0) / \omega_{\rm E} \Omega$ is a dimensionless coupling constant
with $N(0)$ as the density of states per volume at the Fermi energy.
$\tomega_m$ and $\tth_m$, which are the renormalized Matsubara frequency and magnetic field, respectively, are as follows:
\begin{align}
&\tomega_m = \omega_m + \Sigma^\omega_m,
~~~~~~
\tth_m = h + \Sigma^h_m.
\end{align}
$V^\pm_{lm}$ are effective interactions for the even-frequency ($+$) and odd-frequency ($-$) components
with respect to both the $\omega_l$ and $\omega_m$ frequencies.
They are given by
\begin{align}
V^\pm_{lm} = \frac{\omegae^2}{(\omega_l-\omega_m)^2+\omegae^2} \pm \frac{\omegae^2}{(\omega_l+\omega_m)^2+\omegae^2}.
\label{eqn:V}
\end{align}
$f^\pm_m$ used in eqs. (\ref{eqn:gap1}) and (\ref{eqn:gap2}) are defined as the real and imaginary parts of
\begin{align}
\frac{\pi}{ \sqrt{ \left( \tomega_m + \ri\tth_m \right)^2
                 + \left( - \Deltaev_m + \ri \Deltaod_m \right)^2 } }
\equiv f^+_m + \ri f^-_m.
\label{eqn:F}
\end{align}
Here, we adopted a sufficiently large cutoff in the Matsubara summation
and checked that there is no cutoff dependence in the self-consistent solutions.

The two matrix equations in eq. (\ref{eqn:gap}) are coupled via the $f^\pm_m$ functions
that contain the normal and anomalous parts of self-energies.
We note that the superscripts $\pm$ in $f^\pm_m$ represent even and odd functions as in $V^\pm_{lm}$.
Therefore, eq. (\ref{eqn:gap}) clearly shows
that $\Sigma^h_l$ and $\Deltaev_l$ ($\Sigma^\omega_l$ and $\Deltaod_l$) have the even (odd) -frequency dependence.
We note that the relative sign between $\Deltaod_l$ and $\Deltaev_l$ is the same as the sign of $h$.
In the absence of the external field ($h=0$), it is expected that $\tth_m=0$, $f^-_m=0$, and $\Deltaod_m=0$.
In this case, eq. (\ref{eqn:gap}) reduces to its typical form for the even-frequency superconductivity.

\begin{figure}[t]
\begin{center}
\includegraphics[width=7.5cm]{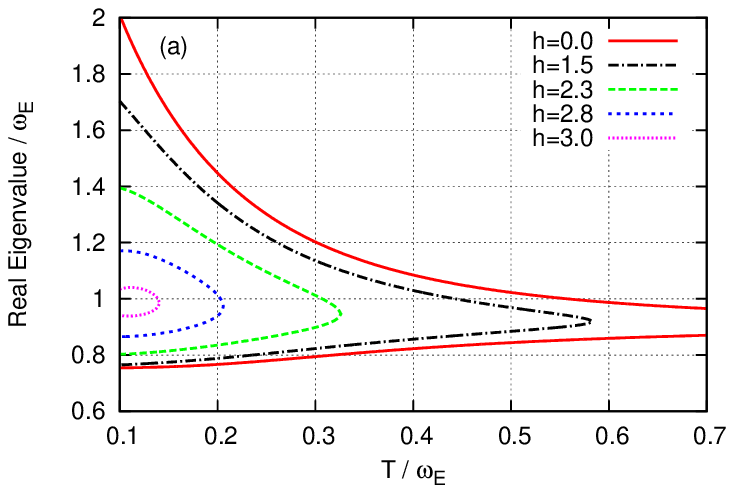}
\includegraphics[width=7.5cm]{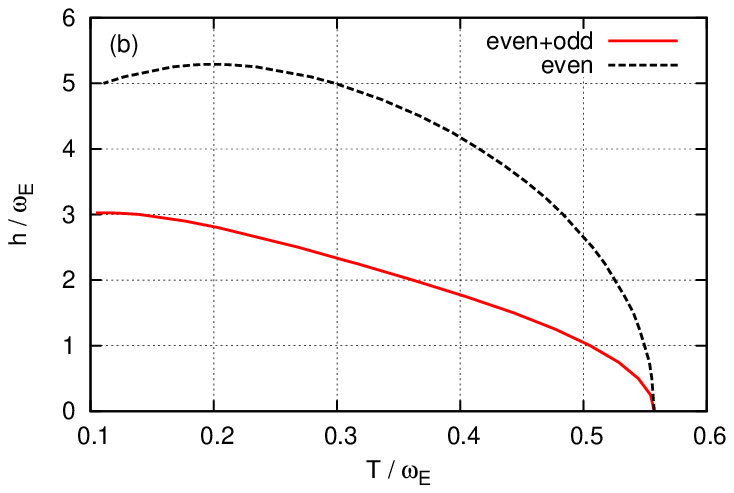}
\end{center}
\caption{
(Color online)
(a) $T$ dependence of the largest two real eigenvalues for $\lambda=10$.
The magnetic fields $h$ are given in units of $\omega_{\rm E}$.
(b) Phase boundary on the $h$-$T$ plane.
The solid line represents the results of the coupled even- and odd-frequency components,
while the dashed line is obtained by retaining the even-frequency component alone.
}
\label{fig:eigen}
\end{figure}

First, we determine $\Tc$ by solving the linearized gap equation.
For this purpose, we put $\Deltaev_m=\Deltaod_m=0$ in eq. (\ref{eqn:F})
and calculate the normal self-energy using eq. (\ref{eqn:gap1}).
With the use of the determined $\tomega_m$ and $\tth_m$, the gap equation eq. (\ref{eqn:gap2}) reduces to an eigenvalue problem.
$\Tc$ is determined when the eigenvalue is equal to 1.
For $h=0$, the matrix for the eigenvalue problem reduces to $\lambda T V^+_{lm}f^+_m$.
In this case, all the eigenvalues are real, since the matrix can be symmetrized.
In contrast to this, the matrix is no longer Hermite for $h\neq 0$ owing to the odd-frequency component
and the eigenvalues are complex in general.
In Fig. \ref{fig:eigen}(a), we show the largest two real eigenvalues under various fields,
where the coupling constant is chosen to be as relatively strong as $\lambda=10$ to emphasize the effects of the odd-frequency component.
For $h\neq 0$, the real eigenvalues exist only in the low-temperature region.
We can see that $\Tc$ decreases with $h$.
In Fig. \ref{fig:eigen}(b), we show the phase boundary on the $T$-$h$ plane.
To see how the odd-frequency component affects the stability,
we also examine $\Tc$ for a pure even-frequency case
by solving eqs. (\ref{eqn:gap1}) and (\ref{eqn:gap2}) self-consistently with $\Deltaod_m=0$.
This is exactly the solution obtained on the basis of the traditional theory only within the even-frequency component.
We note that $\Tc$ is suppressed by the induced odd-frequency component.
Since $T_{c}({\rm even})$ is higher than $T_{c}({\rm even+odd})$,
one may think that the pure even-frequency pairing is more stable and that the odd-frequency component does not appear.
However, the Eliashberg equation (\ref{eqn:gap}) clearly indicates
that the even-frequency component alone cannot be the solution under a finite field.
We emphasize that the appearance of the odd-frequency component is inevitable in the absence of the time-reversal symmetry as $h\neq 0$.

\begin{figure}[t]
\begin{center}
\includegraphics[width=7.5cm]{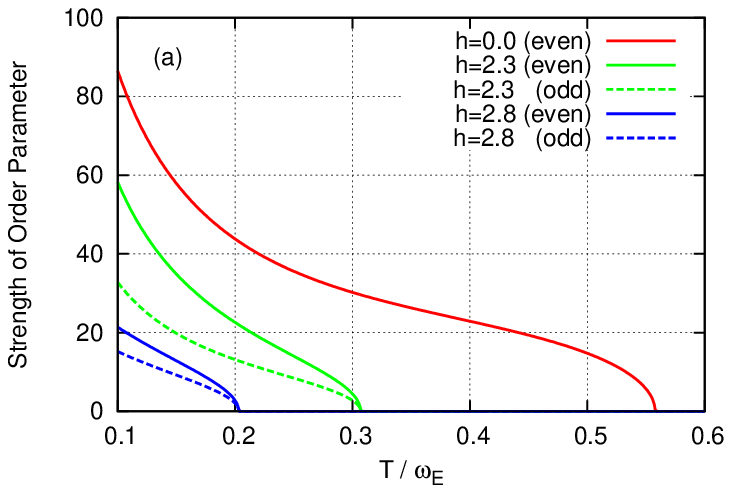}
\includegraphics[width=7.5cm]{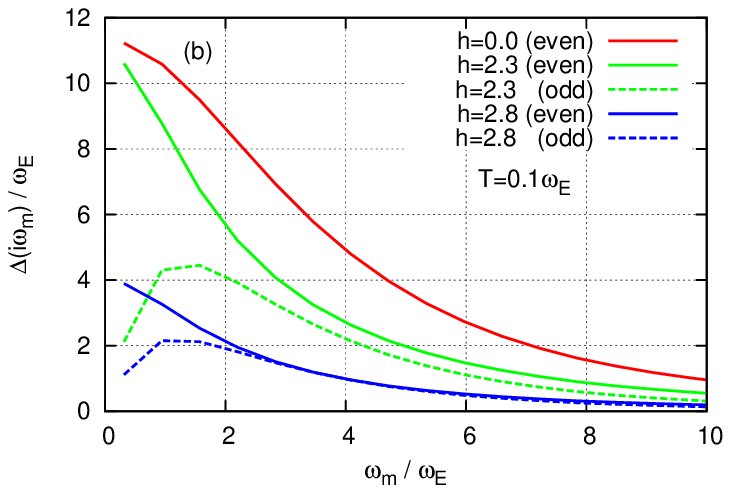}
\end{center}
\caption{
(Color online)
(a) $T$ dependence of the even- and odd-frequency components under various fields for $\lambda=10$.
The strength is defined by
$\sum_{m \ge 0} \Delta(\ri\omega_m) / \omega_{\rm E}$.
(b) $\omega_m$ dependence of the order parameters at $T=0.1 \omegae$.
}
\label{fig:gap}
\end{figure}

Next, we examine the self-consistent solution of the Eliashberg equation.
Figure \ref{fig:gap}(a) shows the temperature dependence of the order parameters
that are summed over the positive Matsubara frequencies to indicate the strength of the frequency-dependent order parameters.
We can see that the odd-frequency component appears below $\Tc$ for $h\neq 0$
and that both components develop at low temperatures.
The odd-frequency component is enhanced by $h$ and the two components become comparable under high fields,
since the off-diagonal matrix element ($f^-_m$) becomes dominant in eq. (\ref{eqn:gap}).
In an extremely high field case, however, inhomogeneous solutions, such as the FFLO state solution, should be taken into account.
\cite{Fulde-FFLO,Larkin}
In contrast to the FFLO state, we emphasize that the effect of the odd-frequency component is present even under low fields.
Figure \ref{fig:gap}(b) shows the frequency dependence of the order parameters for a fixed temperature.
The even-frequency component has a maximum value at the lowest Matsubara frequency,
while the odd-frequency component forms a peak structure at $\omega_m \simeq \omega_{\rm E}$.
This is because the attraction for the odd-frequency component vanishes at low energies, as indicated by eq. (\ref{eqn:V}).

For a weaker coupling constant $\lambda$, $\Tc$ shifts toward the lower temperature region as usual.
The induced odd-frequency component decreases accordingly.
Even in a weak coupling case, however, the odd-frequency component appears below $\Tc$
as long as the magnetic field is finite.
As in Fig. \ref{fig:gap}(a), it is enhanced with the increase in magnetic field.

Finally, we elucidate the stability problem of the gap equation,
which is closely related to the long-standing controversy in the odd-frequency pairing.
That is, it was argued that the pure odd-frequency superconductivity was accompanied by a negative Meissner kernel
\cite{Abrahams-2}
and was thermodynamically unstable in the bulk.
\cite{Heid}
However, the path-integral formulation
\cite{Solenov}
revealed that the odd-frequency superconductivity shows a typical Meissner effect
when the effective free energy is minimized appropriately.
\cite{Belitz-1999,Kusunose-phase}
We discuss this problem in view of gap equation solvability.
For this purpose, we rewrite the anomalous self-energy in a generic form
by introducing the explicit global phase factor $\re^{\ri\varphi}$.
In the coexisting case, the gap equation requires the following form:
\begin{align}
&\bSigma_\Delta(\ri\omega_l) = 
\left(
  \begin{array}{cc}
    0 & \bDelta^{({\rm p})}_l \cr
    \bDelta^{({\rm h})}_l & 0
  \end{array}
\right), \cr
&\bDelta^{({\rm p})}_l = \bDelta_l \re^{\ri\varphi},
~~~~~~
\bDelta^{({\rm h})}_l = - \phi \bDelta^*_l \re^{-\ri\varphi},
\label{eqn:delta-spin} \\
&\bDelta_l =
\left(
  \begin{array}{cc}
    0 & - \Deltaev_l + \ri \Deltaod_l \cr
    \Deltaev_l + \ri \Deltaod_l & 0
  \end{array}
\right).
\nonumber
\end{align}
Here, $\bDelta^{({\rm p})}_l$ and $\bDelta^{({\rm h})}_l$ are the order parameters
for the particle and hole components, respectively.
The relative phase is fixed as `` i " with the real $\Deltaev_l$ and $\Deltaod_l$ as it should be from the structure of the gap equation.
To maintain the common global phase, the factor $\phi$ must be chosen as either $\phi=+1$ or $\phi=-1$.
Then, the denominator in eq. (\ref{eqn:F}) is altered accordingly as
\begin{align}
\sqrt{ \left( \tomega_m + \ri\tth_m \right)^2 + \phi \left( - \Deltaev_m + \ri \Deltaod_m \right)^2 }.
\label{eqn:F2}
\end{align}
It should be emphasized that the solution of the gap equation can be found only when we choose the correct sign of $\phi$.
Namely, from the structure of the gap equation [eq. (\ref{eqn:gap})],
the following condition in eq. (\ref{eqn:F2}) is necessary to obtain the nonzero solution:
\begin{align}
\phi \left( - \Deltaev_m + \ri \Deltaod_m \right)^2>0.
\label{eqn:conditionp}
\end{align}
This is an important clue to understanding the stability problem.

Along this line, let us discuss first the case of a pure even- or odd-frequency superconductivity.
When $h=0$, a pure even-frequency superconductivity is realized.
In this case, the solution of the gap equation can be obtained only when we choose $\phi=+1$ in eq. (\ref{eqn:F2}) as usual.
On the other hand, in the case of a pure odd-frequency superconductivity,
which can be realized by neglecting the mass renormalization in eq. (\ref{eqn:gap}),
\cite{Kusunose-odd}
the solution can be found only when we choose $\phi=-1$ owing to the prefactor `` i " of $\Deltaod_m$ in eq. (\ref{eqn:F2}).
\cite{Kusunose-odd}
When we choose the wrong sign, $\phi=+1$,
no solutions can be obtained for the pure odd-frequency superconductivity.
\cite{linear}
In summary for the pure even- or odd-frequency superconductivity, the solvability condition of the gap equation is represented by
\begin{align}
\left[\bDelta^{({\rm p})}_l\right]^\dagger = \bDelta^{({\rm h})}_l.
\label{eqn:condition}
\end{align}
This condition is consistent with that proposed by the path-integral formulation and the minimization of the free energy.
\cite{Solenov,Belitz-1999,Kusunose-phase}

Next, we discuss the case in which the even- and odd-frequency superconductivities coexist.
In this case, $\Deltaev_l$ ($\Deltaod_l$) is the majority (minority) component.
Here, $\Deltaod_l$ is induced by the magnetic field maintaining the relative phase `` i ".
According to the key condition, eq. (\ref{eqn:conditionp}),
the majority component between $\Deltaev_l$ and $\Deltaod_l$ determines the correct sign of $\phi$.
That is, the even-frequency component is the majority component in the present problem and gives the sign $\phi=+1$.
Indeed, the solution of the gap equation can be found only with $\phi=+1$.
In the coexistence case, eq. (\ref{eqn:condition}) is no longer satisfied.
It is valid only for the pure even- or odd-frequency superconductivity.
\cite{Solenov,Belitz-1999,Kusunose-phase}
Instead, we can generalize the stability and the solvability condition for $\phi$ so as to satisfy eq. (\ref{eqn:conditionp}).
In any cases, we note that the generalized condition (\ref{eqn:conditionp}) is consistent with the minimization of the free energy.
This will be discussed elsewhere.

In summary, we studied the superconducting state under the broken time-reversal symmetry
in conventional phonon-mediated superconductors.
We solved the Eliashberg equation self-consistently with a proper mass renormalization.
It is found that the even- and odd-frequency components of the order parameter coexist
as the consequence of the broken time-reversal symmetry.
The induced odd-frequency component affects more or less various physical quantities,
such as transition temperature, density of states, Meissner effect, and Knight shift.
In the Meissner effect, for instance,
the induced odd-frequency component gives a paramagnetic contribution to suppress the Meissner screening.
This is because the minority component suppresses the majority component
owing to the relative phase factor `` i " between $\Deltaev_m$ and $\Deltaod_m$ in eq. (\ref{eqn:F}).
It should be noted that the paramagnetic Meissner effect never exceeds the typical diamagnetic Meissner effect,
since the former comes from the minority component.
The role of such an odd-frequency component of the order parameter has not been considered in past studies on superconductivity.
It is likely prominent in strong coupling superconductors, such as Pb,
where a considerable enhancement of the induced odd-frequency component could be expected.
It is worth seeking for such effects experimentally under magnetic fields or in ferromagnetic superconductors.

This work is supported by a Grant-in-Aid for Scientific Research C (No. 23540414)
from the Japan Society for the Promotion of Science.
One of the authors (H.K.) is supported by a Grant-in-Aid for Scientific Research on Innovative Areas
``Heavy Electrons" (No. 20102008) of the Ministry of Education, Culture, Sports, Science, and Technology (MEXT), Japan.



\begin{thebibliography}{99}
\bibitem{equal-spin}
  This indicates that the $\bd$-vector component along the magnetic field ($d_z$) is mixed with the singlet pairing
  and that the perpendicular components ($d_x$ and $d_y$) are decoupled.
  Owing to the strong mass renormalization effect, both $S_z=\pm 1$ ($d_x$ and $d_y$) triplet pairings are unstable
  and do not appear even under high magnetic fields.
\bibitem{linear}
  Regardless of $\phi$, the linearized gap equation enables us to determine $T_c$,
  since $\Deltaev_m = \Deltaod_m =0$ is taken in eq. (\ref{eqn:F2}).
  This is one of the reasons for the controversy on the stability condition of eq. (\ref{eqn:condition}).
  Below $T_c$, however, it is vital to choose the correct sign of $\phi$ that satisfies eq. (\ref{eqn:conditionp}).
  Only with the correct choice of $\phi$, we can obtain the nonzero self-consistent solution of the gap equation.
\end{thebibliography}
\end{document}